\documentstyle[prl,aps,epsf,twocolumn]{revtex}
\begin{document}
\title{
Aging in a Chaotic System
}
\author{Eli Barkai\\ 
Department of Chemistry,\\
Massachusetts Institute of Technology,\\
Cambridge, MA 02139.\\
}
\date{\today}
\maketitle

\begin{abstract}

{\bf Abstract}

 We demonstrate aging behavior in a simple non-linear system.
Our model is a chaotic map which generates deterministically
sub-diffusion. Asymptotic behaviors of the diffusion process
are described using aging
continuous time random walks, introduced previously to model
diffusion in glasses.
 
\end{abstract}

Pacs numbers: 05.45.-a, 02.50-r, 05.60-k\\ 

 Aging behavior is found in complex dynamical systems
like spin glasses, glasses, and polymers \cite{Struick}.
 These systems include many
interacting sub-units and are disordered. In this letter we 
will demonstrate that aging can be found also in low dimensional
non-linear systems. Specifically we will show that 
deterministic diffusion generated by one dimensional
maps exhibits aging. And that statistical properties of the
corresponding trajectories can be analyzed using aging 
continuous time random walks (ACTRW), introduced by Monthus and Bouchaud 
\cite{Month} in the context of aging dynamics in glasses
(see also \cite{Rinn} and Ref. therein).

 There exist several methods to investigate aging. One method is to start
a dynamical process at time $t=-t_a$, then at time 
$t=0$ add a small perturbation to the system.
One eventually measures the response at some time $t>0$.
Alternatively one can measure correlation functions between
physical quantities at time $t=0$ and time $t$, after aging
the system in the interval $(-t_a,0)$. We use the latter approach.
A system exhibits aging if its dynamical properties
depend on $t$ and $t_a$ even in the limit when both are long.
Of-course many systems do not exhibit aging, namely when $t > \tau$,
where $\tau$ is a characteristic time scale of the problem, dynamical
properties of the system are independent of the aging time $t_a$.

 In many cases trajectories generated by deterministic systems,
such as low dimensional Hamiltonians or maps, are highly irregular
and noisy.
For an observer these trajectories seem to be generated by
a stochastic; rather than a deterministic mechanism. Hence analysis of 
chaotic trajectories generated deterministically is often based
on random walk concepts \cite{Klafter1,Swin,Klages}.
 It is well known \cite{Schuster} that both
conservative and dissipative 
deterministic systems may generate normal 
$(\alpha=1)$ or anomalous
$(\alpha \ne 1)$ diffusion for a coordinate $x$
\begin{equation}
\langle x^2 \rangle \sim t^{\alpha},
\label{eqC01}
\end{equation}
where the average is over a set of initial conditions (see details below).
The anomalous behavior is due for example
to long trapping events in vicinity
of unstable fixed points \cite{Geisel,Geisel1}  
or stickiness near 
islands 
in phase space \cite{Zaslavsky}. 
Here we shall investigate a deterministic system
which generates sub diffusion behavior $\alpha<1$ and show how rich 
dynamical features emerge when aging is included. More generally, we
demonstrate that aging can be used to probe dynamics in
deterministic systems.

\begin{figure}
\begin{center}
\epsfxsize=70mm
\epsfbox{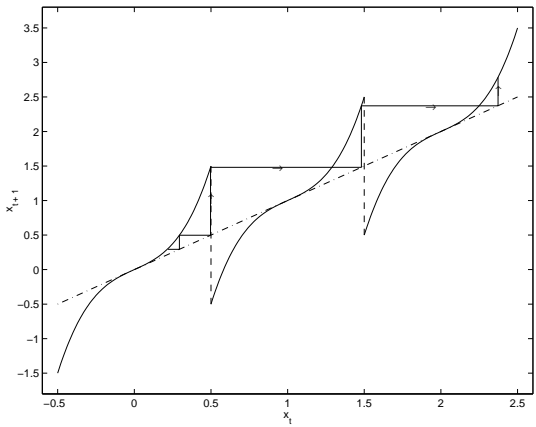}
\end{center}
\caption
{The map $x_{t+1}=x_t + F(x_t)$, defined by Eq. (\protect{\ref{eqC04}}) 
with $z=3$. The linear dash-dot curve is $x_{t+1}=x_t$. The unstable
 fixed points
are on $x_t=0,1,2$.}
\label{fig1}
\end{figure}

 Probably the simplest theoretical tool which generates 
normal and anomalous diffusion are one dimensional maps
\begin{equation}
x_{t+ 1} = x_t + F(x_t)
\label{eqC02}
\end{equation}
with the following symmetry properties of $F(x)$:
$(i)$ $F(x)$ is periodic with a periodicity interval set to
$1$, $F(x) = F(x + N)$, where $N$ is an integer. $(ii)$
$F(x)$ has inversion anti-symmetry; namely,
$F(x) = - F(-x)$.
The study of these maps was motivated by the assumption that
they capture essential features of a driven damped motion
in a periodic potential \cite{Kapral}.
 Geisel and Thomae \cite{Geisel} considered a
rather wide family of such maps which behave 
as
\begin{equation}
F(x) =   a x^z \ \ \mbox{for} \ x \rightarrow + 0,
\label{eqC03}
\end{equation}
where $z>1$. Variations of these maps have been investigated by taking
into account: time dependent noise \cite{Grig},
quenched disorder \cite{Radons}, 
and additional uniform bias which breaks
the symmetry of the map \cite{Bias}.
We use the map
\begin{equation}
F(x) = (2 x)^z, \ \ \ 0\le x \le {1 \over 2}
\label{eqC04}
\end{equation}
which together with the symmetry properties of the map
define the mapping for all $x$. In Fig. \ref{fig1} we show
the map for three unit cells.

 To investigate aging, e.g. numerically, we choose
an initial condition $x_{-t_a}$ which is chosen randomly
and uniformly in the interval $- 1 / 2 < x_{-t_a} < 1 / 2$.
The quantity of interest is the displacement in the interval $(0,t)$,
$x=x_t - x_0$ which is obtained using the map Eq. (\ref{eqC02}).
Previous work \cite{Geisel,ZK}
considered the non-aging regime, namely
$t_a=0$.

 In an ongoing process a walker following the iteration rules
may get stuck close to the vicinity of unstable fixed points
of the map (see Fig. \ref{fig1}). It has been shown, both analytically and
 numerically, that probability density function (PDF) of escape times
of trajectories from the vicinity of the fixed points decays like
a power law \cite{Geisel}.
To see this one considers the dynamics in half
a unit cell, say $0<x<1/2$. Assume that at time
$t=0$ the particle is on $x^{*}$ residing in vicinity of the
the fixed point $x=0$. Close to the fixed point we may approximate the
map Eq. (\ref{eqC02}) with the differential equation
${d x/ dt} = F(x)$. Hence the escape time from $x^{*}$ to
a boundary on $b$ ($x^{*} <b<1/2$) is 
\begin{equation}
t \simeq \int_{x^{*}}^b {d x \over F(x) }
\label{eqCO5}
\end{equation}
using Eq. (\ref{eqC03})
\begin{equation}
t \simeq {1 \over a} \left[ {{x^{*}}^{-z+1} \over z-1} - {b^{-z+1} \over z-1} \right].
\label{eqCO6}
\end{equation}
The PDF of escape times $\psi(t)$ is related to the unknown PDF
of injection points $\eta(x^{*})$, through the chain rule
$\psi(t) = \eta(x^{*} )| d x^{*} / d t|$. Expanding
$\eta(x^{*})$ around the unstable fixed point $x^{*}=0$ 
one finds that for large escape times
\begin{equation}
\psi(t) \sim {A \over \Gamma(-\alpha)}  t^{-1-\alpha},  \ \ \ \  \alpha={1 \over (z-1)},
\label{eqCO7}
\end{equation}
where $A$ depends on the PDF of injection points.
Note that when $z>2$ corresponding to $\alpha<1$, the average escape
time diverges.

To consider stochastic properties of the aging dynamics we
investigate now ACTRW, deriving an explicit expression for the
asymptotic behavior of the Green function.
Specifically we consider a one-dimensional
nearest neighbor random walk. Each lattice point corresponds
to a cell of the iterated maps. Waiting times
on each lattice point are assumed to be described
by $\psi(t)$. Note that after each jumping event we
assume that the process is renewed, namely,  we neglect
correlation between motions in neighboring cells.
This assumption will be justified later using numerical
simulations. As mentioned start of the
ACTRW process is at $t=-t_a$ and our goal is to find
the ACTRW Green function $P(x,t_a,t)$.

 In ACTRW we must introduce the distribution of the first waiting
time $t_1$:  
the time elapsing between start of observation at
 $t=0$ and the first jump event in the interval $(0,t)$.
Let $h_{t_a}(t_1)$ be the PDF 
of $t_1$.
Let $h_s(u)$ be
the double Laplace transform of $h_{t_a}(t_1)$
\begin{equation}
h_s(u) = \int_0^{\infty} {\rm d} t_1 \int_0^{\infty} {\rm d} t_a 
h_{t_a} \left( t_1 \right) e^{- t_a s - t_1 u},
\label{eq03}
\end{equation}
then according to theory of fractal renewal processes \cite{GL}
\begin{equation}
h_s(u) = {1 \over 1 - \psi(s)} {\psi(s) - \psi(u) \over u - s}.
\label{eq04}
\end{equation}
When $\alpha<1$ in Eq. (\ref{eqCO7}) (i.e., $z>2$)
\begin{equation}
h_{t_a}(t_1) \sim { \sin(\pi \alpha) \over \pi} { t_a^{\alpha} \over t_1^{\alpha}
\left( t_1 + t_a \right)},
\label{eq05}
\end{equation}
which is valid in the long aging time limit $t_a >> A^{1/\alpha}$.
Note that Eq. (\ref{eq05}) is independent of the exact form of
$\psi(t)$, besides the exponent $\alpha$. When $\alpha \to 1$
the mass of the PDF $h_{t_a}(t_1 )$ is concentrated in the vicinity
of $t_1 \to 0$, as expected from a `normal process'.
In what follows we will also use the double Laplace transform of Eq.
(\ref{eq05})
\begin{equation}
h_s(u) \sim { u ^{\alpha} - s^{\alpha} \over s^{\alpha} ( u - s)}.
\label{eq06}
\end{equation}

\begin{figure}
\begin{center}
\epsfxsize=70mm
\epsfbox{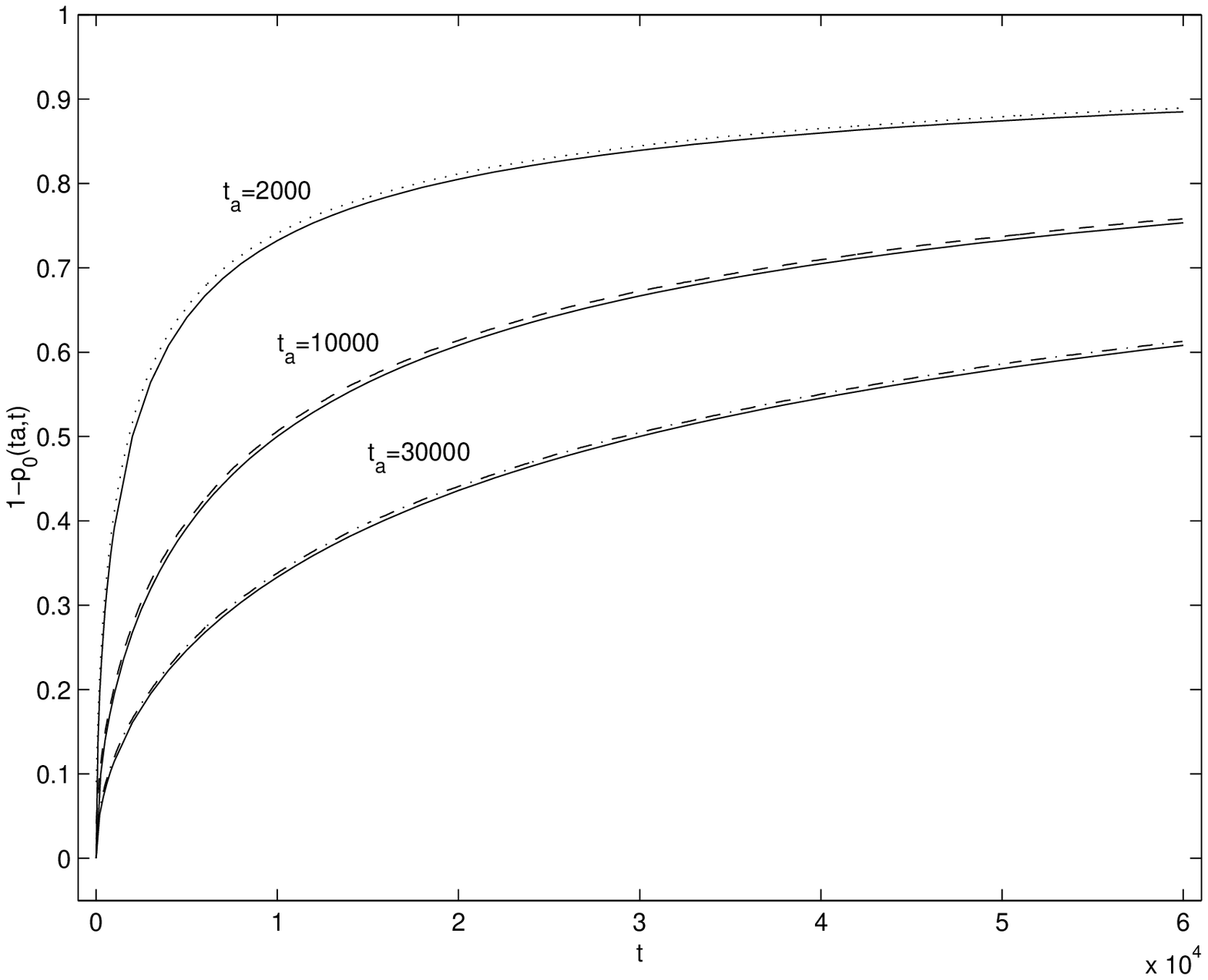}
\end{center}
\caption{
We show the probability of making at least one step in a 
time interval $(0,t)$ for different aging times specified in the figure.
 The solid curve is the theoretical prediction
Eq. (\protect{\ref{eq05}}), the dotted, dashed, and dot dashed curves
are obtained from numerical solution of the map with $z=3$.
}
\label{fig4}
\end{figure}

 We have checked numerically the predictions of Eq. (\ref{eq05})
for $z=3$, analyzing trajectories generated by the
map Eq. (\ref{eqC04}) with three different
aging times. In Fig. \ref{fig4} 
 we show the probability of making  at-least one step
in the interval $(0,t)$:
$\int_0^t h_{t_a}(t) d t \equiv 1 - p_0(t_a,t)$. 
The results show a good agreement between numerical results and
the theoretical prediction Eq. (\ref{eq05}) without fitting.
Fig. \ref{fig4} clearly demonstrates  that as the aging time becomes larger
the time for the first jumping event, from one cell to its neighbor,
becomes larger in statistical sense. 

%
%\begin{figure}[htb]
%\epsfxsize=70mm\baselineskip
%\centerline{\vbox{
%	\epsffile{sprinklerAGE.ps}  }}
%\caption {%\protect\footnotesize
%\label{fig4}
%\end{figure}
%

 We now investigate the ACTRW Green  function.
 Let $p_{n}(t_a,t)$ be the
probability of making $n$ steps in the time interval
$(0,t)$.
 Let $P(k,s,u)$ be the double--Laplace
--Fourier transform $(x \to  k, t_a \to s, t \to u)$ of
$P(x, t_a, t)$, then 
\begin{equation}
P(k,s,u)= \sum_{n = 0}^{\infty} p_n(s,u) \cos^n\left(k\right),
\label{eq07}
\end{equation}
where $p_n(s,u)$ is the double Laplace transform of
$p_n(t_a,t)$. In Eq. (\ref{eq07}) $\cos^n(k)$ is
the characteristic function of a random walk with exactly
$n$ steps. Using convolution theorem of Laplace transform
\begin{equation}
p_n(s,u) = \left\{
\begin{array}{l l}
{1- s h_s(u) \over s u} \ & n=0 \\
\ & \  \\
h_s(u) \psi^{n-1}(u) {1 - \psi(u) \over u} \ & n \ge 1 .
\end{array}
\right.
\label{eq08}
\end{equation}
Hence inserting Eq. (\ref{eq08}) in Eq. (\ref{eq07}), using Eq. (\ref{eq04})
and summing
we find
\begin{equation}
P(k,s,u) = { 1  \over  s u} + {\left[ \psi\left(u\right) - \psi\left(s \right)\right] \left[ 1 - \cos\left( k \right)\right] \over u \left( u -s \right) \left[ 1 - \psi\left(s \right) \right] \left[ 1 - \psi\left(u \right) \cos \left( k \right) \right] }.
\label{eq09}
\end{equation}
We note that only if the underlying process is a Poisson
process, the Green function $P(x,t_a,t)$ is independent
of $t_a$.

 By differentiating Eq. (\ref{eq09}) with respect to
$k$ twice and setting $k=0$, we obtain the mean square displacement
of the random walk
\begin{equation}
\langle x^2 \left( s, u \right) \rangle = 
{ h_s(u) \over u \left[ 1 - \psi\left(u\right) \right] }
\label{eq11}
\end{equation}
From Eq. (\ref{eqCO7}) we have $\psi(u) = 1 - A u^{\alpha} \cdots$,
when $u$ is small.
Using Tauberian theorem one can show that for $t,t_a \gg A^{1/\alpha}$
\begin{equation}
\langle x^2 \left(t_a, t\right) \rangle \sim { 1 \over A} { 1 \over \Gamma\left( 1 + \alpha \right) } \left[ \left( t + t_a \right)^{\alpha} - t_a^{\alpha}\right].
\label{eq12}
\end{equation}
For times $t>>t_a$ we recover standard  
CTRW behavior, $\langle x^2 \left(t_a, t\right) \rangle \propto t^{\alpha}$
\cite{Weiss1}.
For $t/t_a << 1$ we find
$\langle x^2 ( t , t_a) \rangle \propto t/t_a^{1- \alpha}$,
hence as $t_a$ becomes larger the diffusion in this regime
is slowed down. 

\begin{figure}
\begin{center}
\epsfxsize=70mm
\epsfbox{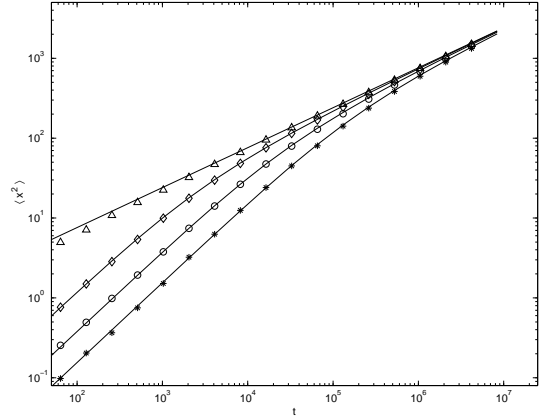}
\end{center}
\caption{
We show the mean square displacement versus the forward time
$t$ for different aging times $t_a$, $t_a=0$ triangle, $t_a=1000$
diamond, $t_a=10000$ circle, and $t_a=60000$ star. The solid curves
are the theoretical prediction Eq. (\protect{\ref{eq12}}).
Here $z=3$.
}
\label{fig3}
\end{figure}

 This  ACTRW behavior is
shown in Fig. \ref{fig3} for the iterated map.
We show $\langle x^2 (t_a,t)\rangle$ versus time
for different values of $t_a$.
Good agreement between ACTRW and the numerical simulations is
found. 
All the curves in Fig. \ref{fig3}
converge into $\langle x^2 (t_a,t)\rangle
\simeq t^{\alpha}$ for long forward times $t$, while at shorter times
the diffusion is clearly slowed down as the aging time is increased.

%

%
%\begin{figure}[htb]
%\epsfxsize=70mm\baselineskip
%\centerline{\vbox{
%	\epsffile{x2age.ps}  }}
%\caption {%\protect\footnotesize
%}
%\label{fig3}
%\end{figure}
%

 To investigate properties of the Green function $P(x,t_a,t)$ 
in the limit of long $t$ and $t_a$ we consider the continuum approximation
of Eq. ({\ref{eq09}) (soon to be briefly justified).
We insert the large wave length expansion
$\cos(k)=1 - k^2/2 $ and low frequency expansion
 $\psi(u) = 1 - A u^{\alpha}$
in Eq. (\ref{eq09}) 
\begin{equation}
P\left( k, s, u \right) \simeq { s^{\alpha} u - s u^{\alpha} \over s^{\alpha + 1} u \left( u - s \right)} + {\left( u^{\alpha} - s^{\alpha} \right) \over s^{\alpha} \left(u - s\right) } {A u^{ \alpha - 1} \over A u^{\alpha} + k^2/2 }.
\label{eq13m1}
\end{equation}
Inverting to the double time $(t,t_a)$
 -- real space $x$  domain
we find that the Green function is a sum of two terms:
 $$ P(x, t_a , t) \sim 
p_{0} \left(t_a , t\right) \delta\left(x\right)  +  $$
\begin{equation}
{ \sin\left( \pi \alpha \right) \over \pi } 
{ 1 \over  t_a \left( { t \over t_a} \right)^{\alpha} \left( 1 + { t \over t_a} \right) } 
\otimes 
{ t |x|^{-(1+2/\alpha)}  \over \alpha \left(2 A\right)^{1/\alpha} } l_{\alpha/2} \left( { t |x|^{-(2/\alpha)}  \over \left( 2 A \right)^{1/\alpha}} \right)
\label{eq13}
\end{equation}
where in this limit
\begin{equation}
p_{0} (t_a , t) \sim { \sin\left( \pi \alpha\right) \over \pi } \int_{t/t_a}^{\infty} { {\rm d} x \over x^{\alpha} \left( 1 + x \right)}.
\label{eq14}
\end{equation}
The first term on the right hand side of Eq. (\ref{eq13})
is a singular term. It corresponds to a random walk
which did not make a jump in the time interval $(0,t)$.
The symbol $\otimes$ in the second term
 in Eq. (\ref{eq13}) is the Laplace convolution operator
with respect to
the forward time $t$,
$l_{\alpha/2}(t)$ is the one sided L\'evy stable PDF, whose
Laplace pair is $\exp( - u^{\alpha/2} )$.
It is easy to see that the asymptotic solution
Eq. (\ref{eq13}) is non negative and normalized [proof:
set $k=0$ in Eq.
(\ref{eq13m1})]. In the limit $\alpha \to 1$ we get
a Gaussian Green function which is independent of $t_a$,
hence in the normal transport regime no aging behavior
is found
[proof: set $\alpha=1$ in Eq. (\ref{eq13m1})].

 To justify the approximation Eq. (\ref{eq13}) we have calculated
the even moments of the ACTRW exactly, in $u,s$ space, using Eq. (\ref{eq09})
[i.e., $\langle x^2(s,u) \rangle$ Eq. (\ref{eq11}), $\langle x^4(s,u) \rangle$,
etc]. 
Then using Tauberian theorems we can show that
for long $t$ and long $t_a$, the ratio $t/t_a$ being arbitrary,
these moments and the moments obtained directly from the
approximation Eq. (\ref{eq13}) are identical. 
In this way we have justified the continuum approximation. 
The details of this calculation
and generalizations to other classes of random walks 
will be published elsewhere. 

 The behavior of the Green function Eq. (\ref{eq13}) 
is shown in Fig.  (\ref{fig5}).
Not shown is 
the behavior on the origin which exhibits a singular behavior [i.e, 
the $\delta(x)$ term in Eq. (\ref{eq13})]. The behavior
of this singular term was displayed already in Fig. \ref{fig4}. 
A good agreement between simulations and the ACTRW Green
function is obtained.
 
Using Eq. (\ref{eq13}) we find the non-singular part of
the ACTRW on the origin
\begin{equation}
P(x,t_a,t)|_{x=0} = t^{ - \alpha/2} g\left( { t \over t_a} \right),
\label{eq15}
\end{equation}
where 
\begin{equation}
g(x) = x^{ \alpha /2} { \sin\left( \pi \alpha \right) \over 2 \pi \Gamma\left( 1 - \alpha/2\right)} \int_0^x {\rm  d} y { \left( x - y \right)^{ - \alpha/ 2} \over \left( 1 + y \right) y^{ \alpha} } .
\label{eq16}
\end{equation}
Hence we find the asymptotic behaviors 
\begin{equation}
P(x,t_a,t)|_{x=0} \sim  \left\{
\begin{array}{l l}
{ t^{ - \alpha/2} \over 2 \Gamma\left( \alpha \right) \Gamma \left( 2 - 3 \alpha /2 \right) } \left( { t \over t_a } \right)^{ 1 - \alpha} \ & t \ll t_a \\
\ & \  \\
{ t^{ - \alpha/2} \over 2 \Gamma\left( 1- \alpha/2 \right) } \ & t \gg t_a .
\end{array}
\right.
\label{eq17}
\end{equation}
In the limit of $t \gg t_a$ we recover the standard
CTRW behavior \cite{Weiss1}.
Note however that the convergence of ACTRW to standard CTRW
behavior may become extremely slow when $\alpha$ is small.
For example when $\alpha=1/12$, 
and  $t/t_a=10^{8}$, large deviations from the asymptotic limit
are found:
$2 \Gamma(1-\alpha/2) t^{\alpha/2} P(x,t_a,t)|_{x=0}=0.788$
 instead of the asymptotic
value $1$. Since the limit $\alpha \to 0$ is important for several
systems \cite{KlafterD}, 
(i.e., logarithmic diffusion as found in Sinai's model),
it is clear that aging can be important
also when $t>t_a$.

\begin{figure}
\begin{center}
\epsfxsize=70mm
\epsfbox{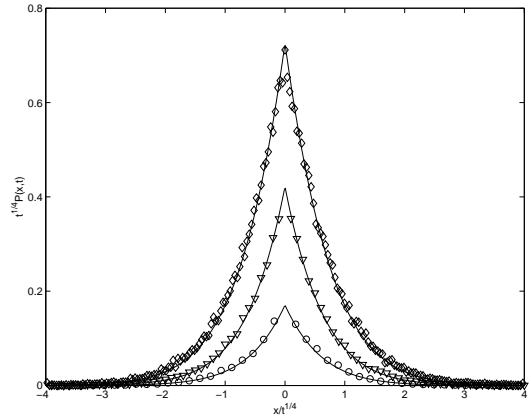}
\end{center}
\caption{
We show the Green function obtained from simulation
of the deterministic map  in {\em scaling form} 
for: (i) $t_a=10^4,t=10^3$, circles,
(ii) $t_a=10^4,t_a=10^4$, triangles, (iii) and $t=4*10^5,t_a=0$, diamonds. 
The curves are
our theoretical results which are  in good agreement with the simulations
with $z=3$.
}
\label{fig5}
\end{figure}
%

%\begin{figure}[htb]
%\epsfxsize=70mm\baselineskip
%\centerline{\vbox{
%	\epsffile{actrw.eps}  }}
%\caption {%\protect\footnotesize
%}
%\label{fig5}
%\end{figure}

 To conclude, we have demonstrated for the first time 
aging behavior in diffusion generated
by iterated maps. In the regime $t \le t_a$ we observe 
a slowing down of the dynamics as the aging time is increased.
In the limit $t >> t_a$ we obtain previous results, the convergence
towards this regime can become extremely slow when $\alpha$ is small.
The aging dynamics of the iterated maps  
exhibits a universal behavior in the sense that the Green function, 
$P(x,t_a,t)$
does not depend
on precise shape of $\psi(t)$ besides the exponent $\alpha=(z-1)^{-1}<1$. 
We believe that aging dynamics can be found also in 
other low dimensional deterministic systems, 
an issue left for future research.

{\bf Acknowledgments} I thank J. P. Bouchaud for pointing out Ref. 
\cite{GL} and Yuan-Chung Cheng for helping with the numerics.


\begin{thebibliography}{99}

\bibitem{Struick} L.C.E. Struick, {\em Physical Aging in Amorphous Polymers
and Other Materials} (Elsevier, Houston, 1978).

\bibitem{Month} C. Monthus and J. P. Bouchaud, {\em J. Phys. A} {\bf 29}, 3847 (1996). 

\bibitem{Rinn} B. Rinn, P. Maass, and J. P. Bouchaud
{\em Phys. Rev. Lett.} {\bf 84}, 5403 (2000). 

\bibitem{Klafter1} J. Klafter, M. F. Shlesinger and G. Zumofen, {\em Phys. Today} {\bf 49} (2) 33 (1996).

\bibitem{Swin} T. H. Solomon, E. R. Weeks, and H. L. Swinney,  {\em Phys. Rev. Lett.} {\bf 71}, 23 (1995). 

\bibitem{Klages} R. Klages, and J. R. Dorfman, {\em Phys. Rev. Lett.} {\bf 74}, 387 (1995).

\bibitem{Schuster} H. G. Schuster,
 {\em Deterministic Chaos} (VCH Verlags-gesellschft mbH, Weinheim 1989). 
E. Ott, {\em Chaos in Dynamical Systems} (Cambridge University Press, Cambridge, 1993).

\bibitem{Geisel} T. Geisel, and S. Thomae, {\em Phys. Rev. Lett.} {\bf 52},
1936 (1984).

\bibitem{Geisel1} T. Geisel, J. Nierwetberg, and A. Zacharel,
{\em Phys. Rev. Lett.} {\bf 54}, 616 (1985).

\bibitem{Zaslavsky} G. M. Zaslavsky, M. Edelman, and B. A. Niyazov,
{\em Chaos} {\bf 7}, 159 (1997).

\bibitem{Kapral} M. Schell, S. Fraser, and R. Kapral, {\em Phys. Rev. A}, {bf 26}, 504 (1982).

\bibitem{Grig} R. Bettin, R. Mannella, B. J. West, and P. Grigolini,
{\em Phys. Rev. E.} {\bf 51}, 212 (1995).

\bibitem{Radons} G. Radons, {\em Phys. Rev. Lett.} {\bf 77}, 23 (1996).

\bibitem{Bias} E. Barkai, and J. Klafter, {\em Phys. Rev. Lett.} {\bf 79}, 2245
(1997).

\bibitem{ZK} G. Zumofen, and J. Klafter, {\em Phys. Rev. E.} {\bf 47}, 851
(19993).

\bibitem{GL}  C. Gordeche, and J. M. Luck, {J. of Statistical Physics} {\bf 104}
489 (2001).

\bibitem{Weiss1} R. Metzler, and J. Klafter, {\em Phys. Rep.} {\bf 339} 1 (2000).

\bibitem{KlafterD} J. Drager, and J. Klafter, {\em Phys. Rev. Lett.} {\bf 84} 5998 (2000).


\end{thebibliography}
\end{document}